\documentclass[aps,showpacs,twocolumn,longbibliography]{revtex4-1}
\usepackage[utf8]{inputenc}
\usepackage{amsfonts}
\usepackage{amssymb}
\usepackage{amsmath}
\usepackage{graphicx}
\usepackage{epsfig}
\usepackage{subfigure}
\usepackage{appendix}
\usepackage{color}
\usepackage{hyperref}
\usepackage{clrscode3e}
\usepackage{fullpage}

\hypersetup{hypertex=true,
colorlinks=true,
linkcolor=blue,
anchorcolor=blue,
urlcolor=blue,
citecolor=blue}
\begin{document}

\title{Formations of generalized Wannier-Stark ladders: Theorem and
applications}
\author{H. P. Zhang}
\author{Z. Song}
\email{songtc@nankai.edu.cn}

\begin{abstract}
The Wannier-Stark ladder (WSL) is a basic concept, supporting periodic
oscillation, widely used in many areas of physics. In this paper, we
investigate the formations of WSL in generalized systems, including strongly
correlated and non-Hermitian systems. We present a theorem on the existence
of WSL for a set of general systems that are translationally symmetric
before the addition of a linear potential. For a non-Hermitian system, the
WSL becomes complex but maintains a real energy level spacing. We illustrate
the theorem using 1D extended Bose-Hubbard models with both real and
imaginary hopping strengths. It is shown that the Bloch-Zener oscillations
of correlated bosons are particularly remarkable under resonant conditions.
Numerical simulations for cases with boson numbers $N=2$, $3$, and $4$ are
presented. Analytical and numerical results for the time evolution of the $N$%
-boson-occupied initial state indicate that all evolved states exhibit quasi
periodic oscillations, but with different profiles, depending on the
Hermiticity and interaction strength.
\end{abstract}

\maketitle

\affiliation{School of Physics, Nankai University, Tianjin 300071, China}

\section{Introduction}

\label{Introduction} Rigorous results in quantum { interacting}
systems are rare but crucial for gaining valuable insights into the
characterization of dynamic behaviors within correlated systems. For
non-interacting system, Wannier-Stark ladders (WSL) are a theoretical model
in solid-state physics that describe the quantization of electronic states
in a crystal under the influence of a constant electric field \cite%
{Bloch1929,wannier1959elements,Wannier1960,Glueck2002,Waschke1993}. This has
been observed in systems of semiconductor superlattices and ultracold atoms 
\cite{Waschke1993,BenDahan1996,Wilkinson1996,Anderson1998}. It is closely
related to Bloch oscillations (BOs) due to the equidistant spectrum. In the
past decade, such a phenomenon has attracted much attention in cold-atoms
physics and photonics due to applications in interferometric measurements
and as a method for manipulating localized wave packets \cite%
{Breid2006,Breid2007,Dreisow2009,Kling2010,Ploetz2011}. It can be simulated
using artificial quantum systems, such as superconducting circuits \cite%
{Song2024}. On the other hand, the dynamics of particle pairs in lattice
systems have garnered considerable interest, owing to the rapid advancements
in experimental techniques. Ultra-cold atoms have proven to be an ideal
testing ground for few-particle fundamental physics, as optical lattices
offer clean realizations of a variety of { interacting Hamiltonian
systems}. It stimulates many experimental \cite%
{Winkler2006,Foelling2007,Gustavsson2008} and theoretical investigations 
\cite%
{Mahajan2006,Petrosyan2007,Creffield2007,Kuklov2007,Zoellner2008,Wang2008,Valiente2008,Jin2009,Valiente2009,Valiente2010,Javanainen2010,Wang2010,Rosch2008,Zhang2024}
in strongly correlated systems. Recently, it has been shown that a bound
pair within a correlated system also exhibits periodic dynamics when
subjected to a linear potential \cite%
{Khomeriki2010,Longhi2011,Longhi2012,Corrielli2013,Lin2014,ZhangXZ2016,Zhang2024}%
.

In this work, we extend the concept of WSL to the { interacting
Hamiltonian systems}. We investigate the formations of WSL in generalized
systems, including strongly correlated and non-Hermitian systems. We present
a theorem on the existence of WSL for a set of general systems that are
translationally symmetric before the addition of a linear potential. We show
that the WSL becomes complex but maintains a real energy level spacing for a
non-Hermitian system. To illustrate the theorem, we focus on 1D extended
Bose-Hubbard models with both real and imaginary hopping strengths. It is
shown that the Bloch-Zener oscillations of correlated bosons are
particularly remarkable under resonant conditions. Numerical simulations for
cases with boson numbers $N=2$, $3$, and $4$ are presented. Analytical and
numerical results for the time evolution of the $N$-boson-occupied initial
state indicate that all evolved states exhibit quasi periodic oscillations,
but with different profiles, depending on the Hermiticity and interaction
strength.

This paper is organized as follows. In Sec. \ref{Theorem on WSLs}, we
present the theorem on the structure of energy levels for general systems in
a linear field, including Hermitian and non-Hermitian. { In Sec. \ref{Extended
Bose-Hubbard model}, we take the extended Bose-Hubbard model as an
application of the theorem and provide its effective Hamiltonian 
under the resonant condition in strongly interaction limit. Sec. \ref%
{Dynamics of effective Hamiltonian} dedicated to the dynamics of the 
few-boson effective Hamiltonian and the corresponding original Hamiltonian,
respectively. In particular, we also provide numerical simulations
for both Hermitian and non-Hermitian sample systems, which offer evident
demonstrations of our rigorous results, in Sec. \ref{Dynamics of correlated
bosons}.} Finally, we provide a summary in Sec. \ref{Summary}.

\section{Theorem on WSLs}

\label{Theorem on WSLs}

In general, a linear field can break the translational symmetry and disrupt
the energy band structure of a non-interacting system. For some simple
systems, it has been precisely demonstrated that the formation of WSLs leads
to periodic dynamics, specifically Bloch oscillations. However, obtaining
the energy levels for complex systems, particularly those with interactions,
is quite challenging. The main aim of this paper is to determine whether
WSLs still exist in general systems. In this section, we present a theorem
on the structure of energy levels for general systems. We focus solely on
the 1D system for simplicity. The obtained conclusion is applicable to
higher-dimensional systems.

We start with a general tight-binding model with the Hamiltonian in the form

\begin{equation}
H=H_{0}+\omega\sum_{j=-\infty }^{\infty }\sum_{\sigma =1}^{\Lambda
}ja_{j,\sigma }^{\dag }a_{j,\sigma },
\end{equation}%
where $a_{j,\sigma }^{\dag }$\ ($a_{j,\sigma }$) is the boson or fermion
creation (annihilation) operator, with internal degree of freedom (or
flavor) $\sigma =\left[ 1,\Lambda \right] $, at the $i$th site.

\textit{Theorem.}---The energy levels of Hamiltonian $H$ must consist of
multi-set of WSLs with an identical real level spacing, which is independent
of $H_{0}$, if $H_{0}$ is constructed under the following conditions:

(i) The total particle number $\sum_{j=-\infty }^{\infty }\sum_{\sigma
=1}^{\Lambda }a_{j,\sigma }^{\dag }a_{j,\sigma }$\ is conservative, that is%
\begin{equation}
\lbrack \sum_{j=-\infty }^{\infty }\sum_{\sigma =1}^{\Lambda }a_{j,\sigma
}^{\dag }a_{j,\sigma },H_{0}]=0.
\end{equation}

(ii) Hamiltonian $H_{0}$\ has translational symmetry, that is%
\begin{equation}
\left[ T_{r},H_{0}\right] =0,
\end{equation}%
where $T_{r}$\ is the translational operator defined as 
\begin{equation}
T_{r}a_{j,\sigma }T_{r}^{-1}=a_{j+r,\sigma}.
\end{equation}

\textit{Proof.}--Based on the assumptions on $H$, we simply have%
\begin{equation}
\left[ T_{r},H\right] =-r\omega \sum_{j=-\infty }^{\infty }\sum_{\sigma
=1}^{\Lambda }a_{j,\sigma }^{\dag }a_{j,\sigma }T_{r},
\end{equation}%
which is referred to as a ramped translational symmetry. in the invariant
subspace with fixed particle number $N$, any eigenstate of $H$\ obeys%
\begin{equation}
H|\psi _{0}\rangle =E_{0}|\psi _{0}\rangle ,
\end{equation}%
and%
\begin{equation}
\sum_{j=-\infty }^{\infty }\sum_{\sigma =1}^{\Lambda }a_{j,\sigma }^{\dag
}a_{j,\sigma }|\psi _{0}\rangle =N|\psi _{0}\rangle .
\end{equation}%
Then we always have 
\begin{equation}
H\left( T_{r}|\psi _{0}\rangle \right) =\left( E_{0}+Nr\mathcal{\omega }%
\right) \left( T_{r}|\psi _{0}\rangle \right) ,
\end{equation}%
and 
\begin{equation}
H\left( T_{r}^{-1}|\psi _{0}\rangle \right) =\left( E_{0}-Nr\mathcal{\omega }%
\right) \left( T_{r}^{-1}|\psi _{0}\rangle \right) ,
\end{equation}%
i.e., $T_{r}|\psi _{0}\rangle $ ($T_{r}^{-1}|\psi _{0}\rangle $)\ is also
the eigenstate of $H$\ with eigen energy $E_{0}+Nr\mathcal{\omega }$ ($%
E_{0}-Nr\mathcal{\omega }$). Operator $T_{r}$\ increments the energy level
by a step of $Nr\mathcal{\omega }$ and the operator $T_{r}^{-1}$ decrements
the energy level by a step of $Nr\mathcal{\omega }$.\ We can then construct
a set of eigenstates 
\begin{equation}
|\psi _{m}\rangle =\left( T_{r}\right) ^{m}|\psi _{0}\rangle ,
\end{equation}%
$(m=0,\pm 1,\pm 2,...)$ with eigenenergy 
\begin{equation}
E_{m}=E_{0}+mNr\mathcal{\omega }.
\end{equation}%
It indicates that the translational operator $T_{r}$\ acts as a ladder
operator. This proof is independent of $H_{0}$. We note that based on
another eigenstate $|\psi _{0}^{\prime }\rangle $, which does not be long to 
$\left\{ |\psi _{m}\rangle \right\} $, another set of eigenstates $\left\{
|\psi _{m}^{\prime }\rangle \right\} $ can be generated accordingly.
Although $E_{0}$\ can be real or complex, dependenting on the detail of the
Hamiltonian $H_{0}$, we can conclude that the energy levels of Hamiltonian $%
H $ consists of multi-set of WSLs with an identical real level spacing $Nr%
\mathcal{\omega}$. { In principle, this conclusion is true for any
given $N$. However, a large energy level spacing that
exceeds the scale of low-energy physics is meaningless. 

{Fortunately, this conclusion drawn from few-body systems is applicable to specific invariant subspaces of many-body systems, in which energy eigenstates are localized.} Here, we provide an example to demonstrate this point. Consider a
system with a set of $n$-boson cluster eigenstates in the $n$%
-boson invariant subspace. These eigenstates are connected by the
translational operator. In this situation, one could easily have a large
number $K$ of widely separated $n$-boson clusters to
construct the eigenstates with a total boson number $N=Kn$. The WSL
frequency spacing derived above for states of that multi-cluster kind would
therefore be $Knr\omega$. However, there exist many subsets of
eigenstates, which can be constructed by translating one $n$-boson
cluster while leaving the other $K-1$ clusters unchanged. Then,
the corresponding energy levels in a subset form the WSL with frequency
spacing $nr\omega$ rather than $Nr\omega$. This indicates
that a system with a large boson number $N=Kn$ can also support
dynamics with a smaller frequency $nr\omega$.}

Applying the conclusion to the several cases, the same result can be
obtained in an exact manner. (i) Considering the fermonic Hubbard model,
where the flavor $\Lambda =2$, with $a_{j,1}=c_{j,\uparrow }$\ and $%
a_{j,2}=c_{j,\downarrow }$ being the fermion operators, the doublon Bloch
oscillation is the direct demonstration of the existence of WSLs \cite%
{Longhi2012,Zhang2024}. (ii) Considering the extended Bose-Hubbard model,
where the flavor $\Lambda =1$, with $a_{j,1}=b_{j}$ being boson operator,
the bound pair Bloch oscillation is also the direct demonstration of the
existence of WSLs \cite{Lin2014,ZhangXZ2016}.

This rigorous conclusion has important implications for investigating the
dynamics of { interacting} and non-Hermitian systems in the presence
of a linear field. In the following sections, we will focus on both the
Hermitian and non-Hermitian extended Bose-Hubbard models to demonstrate the
application of our conclusions and to reveal the dynamics of correlated
bosons.

\section{Extended Bose-Hubbard model}

\label{Extended Bose-Hubbard model}

We consider an extended boson Hubbard model describing interacting particles
in the lowest Bloch band of a one dimensional lattice, which can be employed
to describe ultracold atoms or molecules with magnetic or electric
dipole-dipole interactions in optical lattices. We focus on the dynamics of
the bosonic cluster, which is $N$ identical bosons in a bound state. For the
simplest case with $N=2$, it has been demonstrated that, as another type of
bound pair, it allows for correlated single-particle tunneling, as shown in
previous work \cite{Jin2011,Lin2014}. Such a bound pair can act as a
quasi-particle, with an energy band width of the same order as that of a
single particle. One of the objectives of this paper is to demonstrate that
this type of bound state can be generalized to an $N$-boson system.

We consider an illstrative example that $H_{0}$\ is a one-dimensional
extended Bose-Hubbard model with the Hamiltonian%
\begin{eqnarray}
H_{\text{\textrm{B}}} &=&H_{0}+\omega \sum_{j=-\infty }^{\infty }jn_{j}, 
\notag \\
H_{0} &=&-\kappa \sum_{j=1}\left( b_{j}^{\dagger }b_{j+1}+\text{ \textrm{H.c}%
.}\right)  \notag \\
&&+\sum_{j=-\infty }^{\infty }[\frac{U}{2}n_{j}\left( n_{j}-1\right)
+Vn_{j}n_{j+1}],  \label{H_B}
\end{eqnarray}%
where $b_{j}^{\dag }$ and $n_{j}=b_{j}^{\dag }b_{j}$\ is the creation and
number operators of the boson at the $j$th site. Parameters $\kappa $, $U$\
and $V$,\ denote the tunneling strength, on-site and NN interactions between
bosons. In this work, we consider the cases with real and imaginary $\kappa $%
, respectively. Obviously, Hamiltonian $H_{0}$\ satisfies the conditions in
the theorem, and then possesses multi-set of WSLs.

Although we cannot obtain the value of $E_{0}$\ for each set of WSL,
additional some features of the energy levels can be revealed for the
systems under some conditions. In the following, we will consider two
typical cases.

\subsection{$\mathcal{PT}$ symmetry}

We introduce two operators. One is a linear operator $\mathcal{P}$, which
represents the local gauge transformation 
\begin{equation}
\mathcal{P}b_{l}\mathcal{P}^{-1}=(-1)^{l}b_{l}.
\end{equation}%
The other is an anti-linear operator $\mathcal{T}$, which is the
time-reversal operator defined as 
\begin{equation}
\mathcal{T}\sqrt{-1}\mathcal{T}^{-1}=-\sqrt{-1}.
\end{equation}%
We note that the extended Bose-Hubbard $H_{\text{\textrm{B}}}$ has $\mathcal{%
PT}$ symmetry, that is%
\begin{equation}
\mathcal{PT}H_{\text{\textrm{B}}}\mathcal{(PT)}^{-1}=H_{\text{\textrm{B}}},
\end{equation}%
when we consider the imaginary hopping strength, $\kappa =i\left\vert \kappa
\right\vert $. According to non-Hermitian quantum mechanics \cite%
{Scholtz1992}, the Hamiltonian $H_{\text{\textrm{B}}}$ is pseudo-Hermitian,
in which complex energy levels appear in the conjugate pairs. This provides
a constraint on the complex WSLs in the Hamiltonian $H_{\text{\textrm{B}}}$,
which will be utilized in the investigation of dynamics in the following
section.\ 

\subsection{Resonant interactions}

Now we consider the Hamiltonian $H_{\text{\textrm{B}}}$\ under the resonant
condition $U=V\gg \left\vert \kappa \right\vert $. Let us begin by analyzing
in detail the many-boson problem within the Hamiltonian $H_{0}$ with zero $%
\kappa $. It is clear that, in the invariant subspace with fixed boson
number $N$, there is a set of degenerate eigenstates $\left\{ \left\vert
l\right\rangle ,l\in \left( -N\infty ,N\infty \right) \right\} $ of $%
H_{0}\left( \kappa =0\right) $ with eigen energy $\varepsilon (N)$ $%
=UN(N-1)/2$, which are expressed as the form%
\begin{equation}
\left\vert jN+\lambda \right\rangle =\frac{\left( b_{j}^{\dagger }\right)
^{N-\lambda }\left( b_{j+1}^{\dagger }\right) ^{\lambda}}{\sqrt{\left(
N-\lambda \right) !\left( \lambda \right) !}}|\mathrm{vac}\rangle ,
\end{equation}%
with $j\in \left( -\infty ,\infty \right) $\ and $\lambda \in \lbrack 0,N-1]$%
. Here $\left\vert \mathrm{vac}\right\rangle $\ is the vacuum state for the
boson operator $b_{j}$. It is easy to check that the nearest energy level
next to $\varepsilon (N)$\ is $\varepsilon (N)-U$, resulting in an energy
gap of $U$.\ This ensures us to perform the first-order perturbation
approximation when the hopping term\ is considered under the strong
correlation condition where $U\gg \left\vert \kappa \right\vert $.

\begin{figure}[tbp]
\centering
\includegraphics[width=0.9\linewidth]{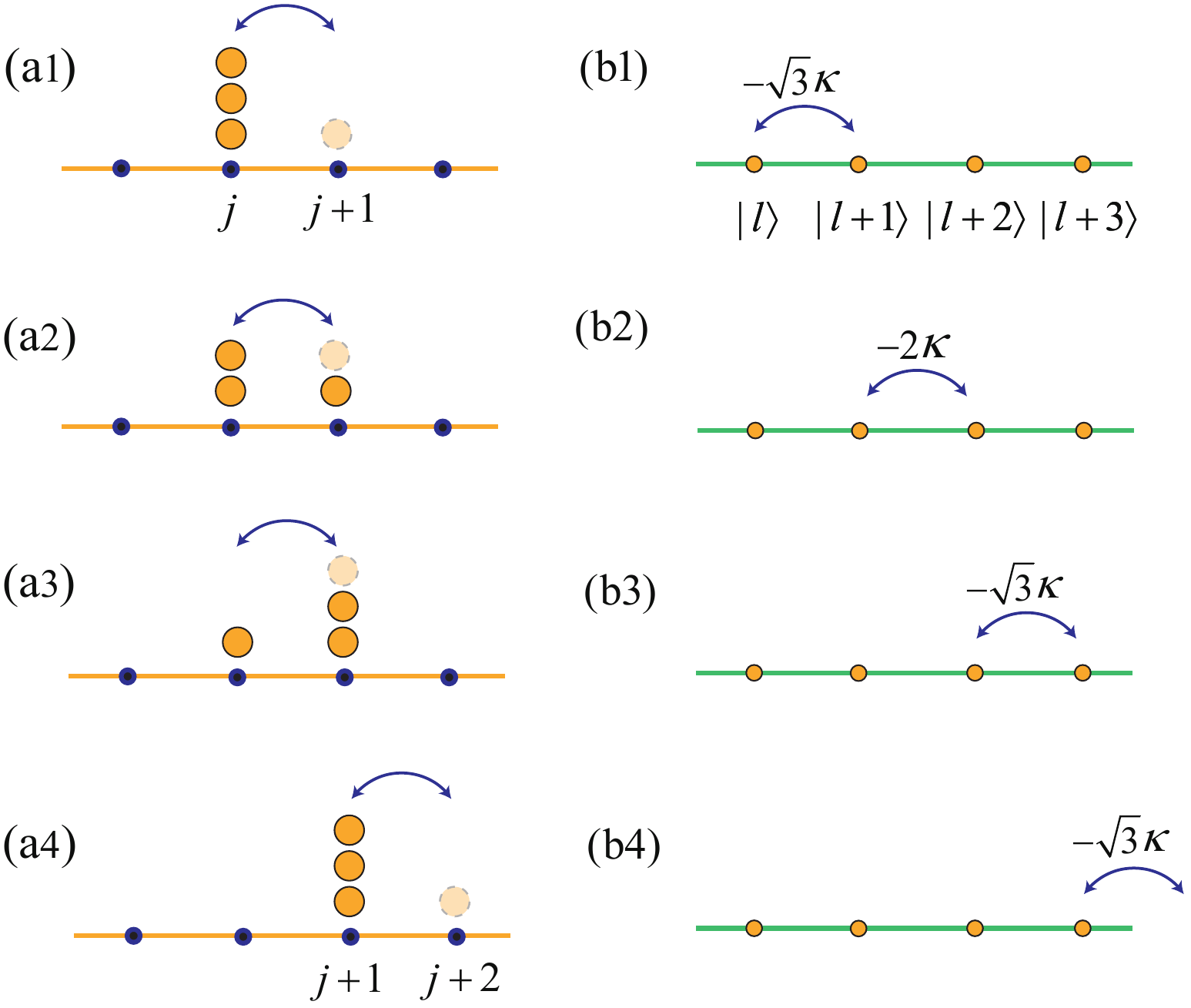}
\caption{ {Schematic illustrations of the correspondence between the
configurations of the boson cluster with }$N=3${\ and the basis
states for the effective Hamiltonian given in Eq.\ \protect\ref{H_eff}.
(a1-a4) Four configurations of the boson cluster with }$N=3${. These
are degenerate states of the Hamiltonian }$H_{0}(\protect\kappa =0)${%
. Here, only the hopping processes within a dimer are presented. (b1-b4) The
sites in the chain represent the basis states of the effective Hamiltonian.
The hopping strengths indicated are non-uniform.}}
\label{slinky_motion_fig}
\end{figure}

In order to obtain the effective Hamiltonian, one can simply evaluate the
nonzero matrix elements of $H_{\text{\textrm{B}}}$\ in the subspace spanned
by the set of eigenstates $\left\{ \left\vert l\right\rangle \right\} $. For
a given boson number $N$, the effective Hamiltonian can be written as%
\begin{eqnarray}
H_{\mathrm{eff}}^{[N]} &=&-\kappa \sum_{j,\lambda }h_{j,\lambda }+U\frac{%
N(N-1)}{2}\sum_{l=-\infty }^{\infty }\left\vert l\right\rangle \left\langle
l\right\vert  \notag \\
&&+\omega \sum_{l=-\infty }^{\infty }l\left\vert l\right\rangle \left\langle
l\right\vert ,  \label{H_eff}
\end{eqnarray}%
where 
\begin{eqnarray}
h_{j,\lambda } &=&\sqrt{(N-\lambda )(\lambda +1)}\left\vert jN+\lambda
\right\rangle  \notag \\
&&\times \left\langle jN+\lambda +1\right\vert +\text{ \textrm{H.c}.}
\end{eqnarray}%
In its present form, $H_{\mathrm{eff}}^{[N]}$ are formally analogous to a
tight-binding model describing a single-particle dynamics in an infinite
chain with NN hopping, under a linear field. We note that the states $%
\left\vert l\right\rangle $\ in each unit cell represent an $N$%
-bosonic-cluster state, in which all $N$ bosons occupy a single dimer of the
original lattice of the Bose-Hubbard system $H_{\text{\textrm{B}}}$. In this
sense, the dynamics of $H_{\mathrm{eff}}^{[N]}$ is essentially\ the dynamics
of $N$ correlated bosons. { In Fig.\ \ref{slinky_motion_fig}, 
a schematic illustration showing the correspondence between the
configurations of the boson cluster with $N=3$ and the basis
states for the effective Hamiltonian under strong interactions is presented.}

Notably, {we would like to point out that the hopping strength in }$H_{%
\mathrm{eff}}^{[N]}${\ is of the order of }$\kappa ${, not }$\kappa ^{2}/U${%
. This ensures that the phenomena arising from the effective Hamiltonian can
be observed in experiments, similar to those of a single boson.}\ On the
other hand, our theorem also holds for the effective Hamiltonian $H_{\mathrm{%
eff}}^{[N]}$, which describes a single-particle system of $N$-site unit
cell. The corresponding step for each ladder is $N\omega $.

\begin{figure*}[tbp]
\centering
\includegraphics[width=0.9\linewidth]{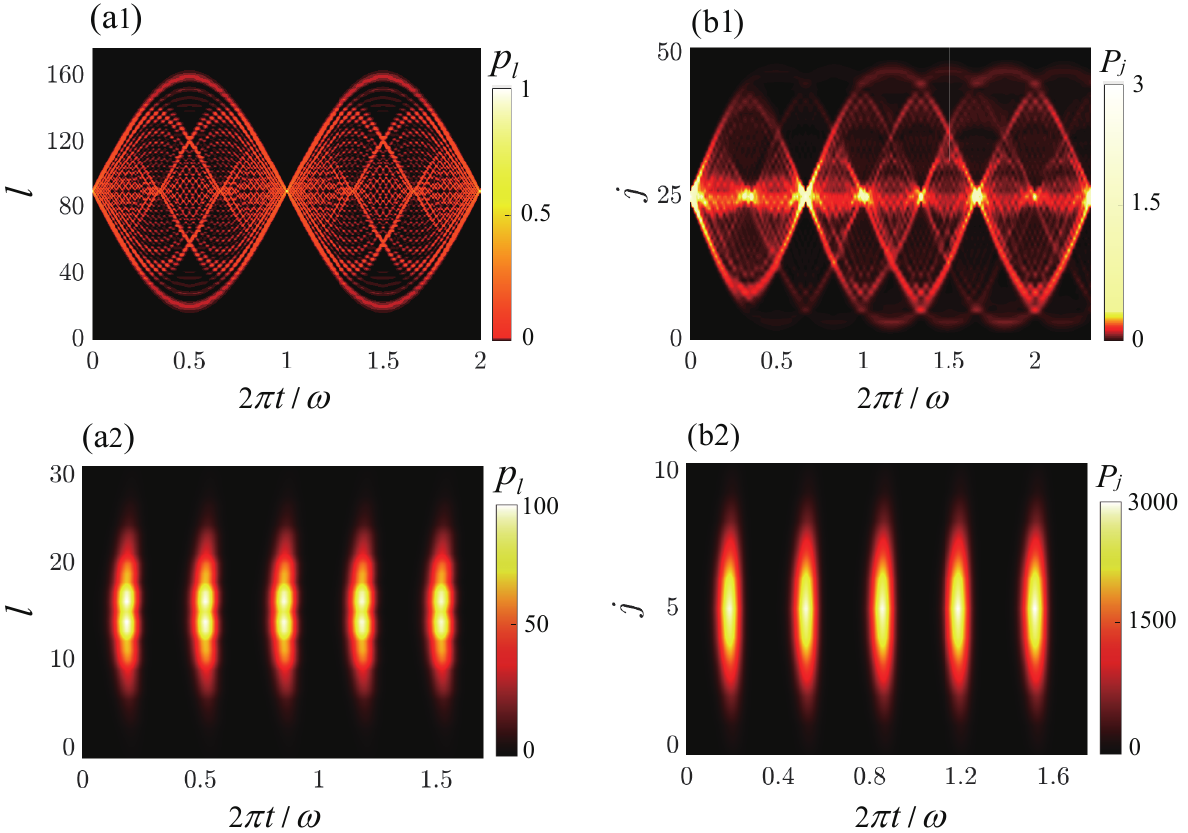}
\caption{Plots of $p_{l}(t)$ defined in Eq.\ (\protect\ref{p_l}) and $%
P_{j}(t)$ defined in Eq.\ (\protect\ref{Pj(t)}), obtained by numerical
diagonalizations. Here, (a1) and (a2) correspond to the effective
Hamiltonian $H_{\mathrm{eff}}^{[3]}$\ with $\protect\kappa =1.0$ and $1.0i$,
respectively. (b1) and (b2) correspond to the Hamiltonian $H_{\mathrm{B}}$\
with $\protect\kappa =1.0$ and $1.0i$, respectively. Other system paraneters
are $\protect\omega =0.1$, $U=V=5.0$, and $N=3$. It indicates that $H_{%
\mathrm{eff}}^{[3]}$\ and $H_{\mathrm{B}}$ have the similar dynamics,
demonstrating a multi-ladder energy level structure.}
\label{fig:fig1}
\end{figure*}
\begin{figure*}[tbp]
\centering
\includegraphics[width=0.9\linewidth]{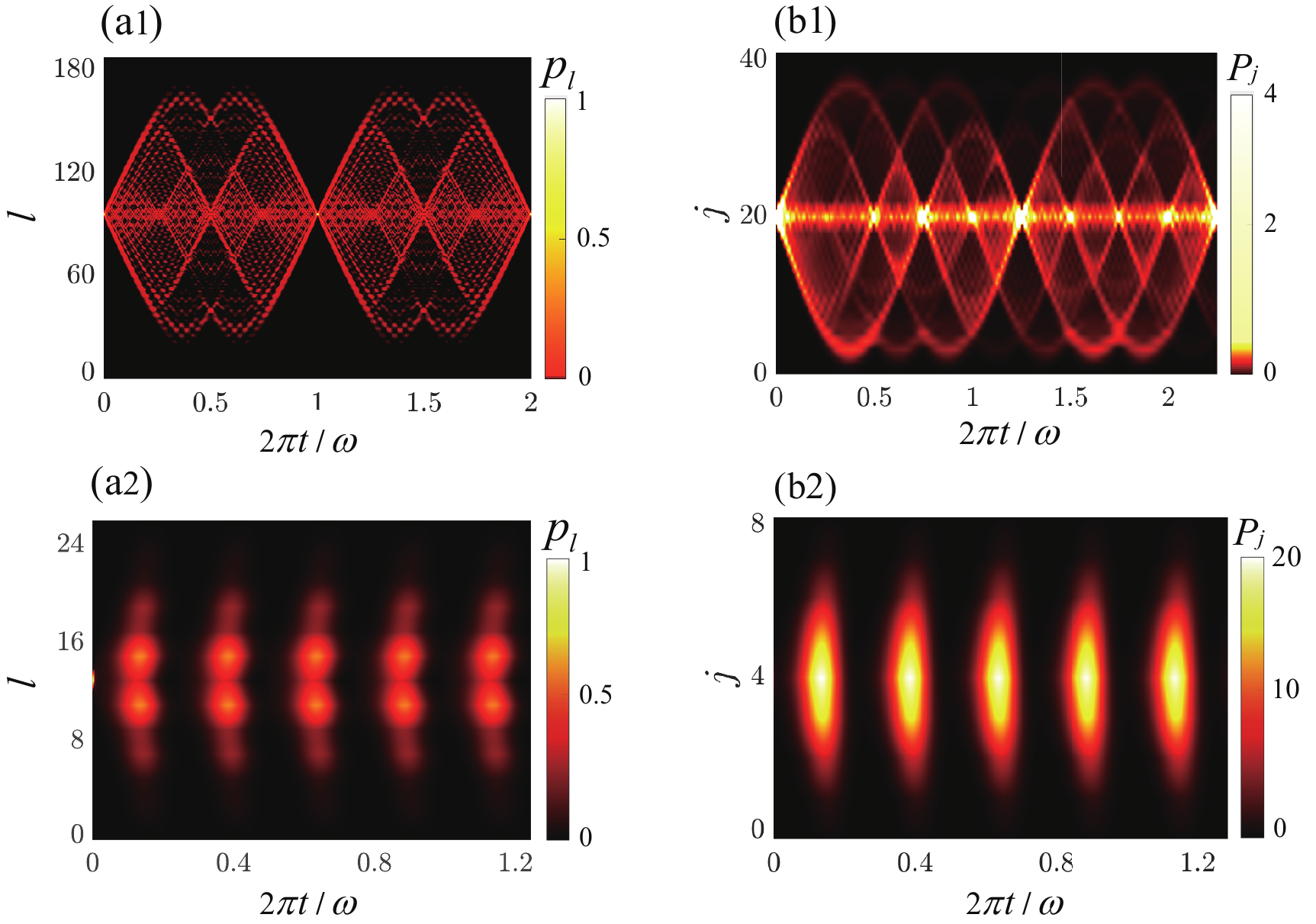}
\caption{The same plots as Fig. \protect\ref{fig:fig1}\ for the case with $%
N=4$.}
\label{fig:fig2}
\end{figure*}
\begin{figure*}[tbp]
\centering
\includegraphics[width=0.85\linewidth]{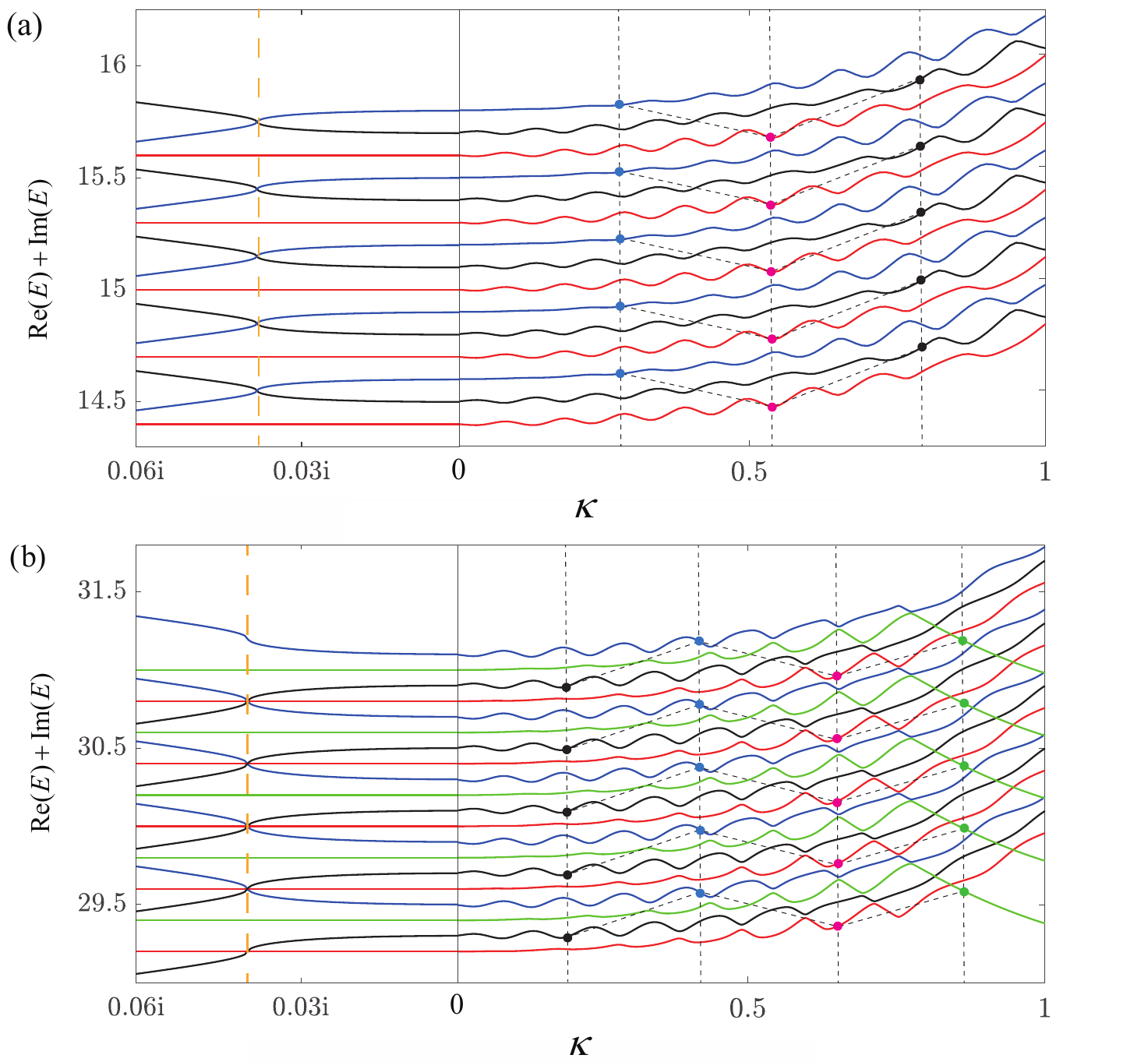}
\caption{Plots of the dressed energy $\protect\varepsilon $\ as a function
of $\protect\kappa $, given by Eq. (\protect\ref{dressed energy}) for
understanding the energy level structures of the extended Bose-Hubbard model
subjected to a linear potential. Several representative energy levels for
the systems are selected: (a) $N=3$ and (b) $N=4$. The system parameters are 
$U=V=5.0$ and $\protect\omega =0.1$. The black dashed lines are added to
indicate the relations between the curves. The orange dashed lines are added
to indicate the coalescing WSL. We see that the patterns consist of three
and four families of curves, respectively.}
\label{fig:fig3}
\end{figure*}

\section{Dynamics of effective Hamiltonian $H_{\mathrm{eff}}^{[N]}$}

\label{Dynamics of effective Hamiltonian}

In this section, we demonstrate the application of the theorem to the
effective Hamiltonian $H_{\mathrm{eff}}^{[N]}$\ with small $N$. We will
present the explicit form of $H_{\mathrm{eff}}^{[N]}$, analyze the relative
energy structures among the sets of ladders based on the symmetry of the
system, and then perform numerical simulations to study the corresponding
dynamics.

Introducing a reflection operator $\mathcal{R}$, which is a unitary operator
and defined as%
\begin{equation}
\mathcal{R}\left\vert l\right\rangle =(-1)^{l}\left\vert -l\right\rangle
\end{equation}%
we have%
\begin{equation}
\mathcal{R}H_{\mathrm{eff}}^{[N]}\mathcal{R}^{-1}=-H_{\mathrm{eff}}^{[N]}.
\end{equation}%
It indicates that the spectrum is symmetric about zero energy because a
unitary transformation preserves the spectrum of the system. This conclusion
holds for the system with an arbitrary value of $\kappa $, including
imaginary numbers, which is associated with complex energy levels. In
addition, $H_{\mathrm{eff}}^{[N]}$\ still possesses $\mathcal{PT}$\
symmetry, given by the relation $\mathcal{PT}H_{\mathrm{eff}}^{[N]}\mathcal{%
(PT)}^{-1}=H_{\mathrm{eff}}^{[N]}$. This ensures that Hamiltonian $H_{ 
\mathrm{eff}}^{[N]}$ is pseudo-Hermitian, meaning that complex energy levels
appear in the conjugate pairs,\ according to non-Hermitian quantum mechanics 
\cite{Scholtz1992}. Therefore, the energy levels of $H_{\mathrm{eff}}^{[N]}$
must satisfy the aforementioned constraints.

In general, the number of the sets of energy ladders is equal to $N$, or the
site number of the unit cell. This feature should be clearly demonstrated in
the dynamics by the occurrence of normal Bloch oscillations and Bloch-Zener
oscillations. In the following, we will detail the level structures for both
the Hermitian and non-Hermitian Hamiltonians $H_{\mathrm{eff}}^{[N]}$,
focusing on small values of $N$.

(i) In the case where $N=2$, the effective Hamiltonian is given by%
\begin{equation}
H_{\mathrm{eff}}^{[2]}=-\sqrt{2}\kappa \sum_{l=-\infty }^{\infty }(|l\rangle
\langle l+1|+\mathrm{H.c.})+\omega\sum_{l=-\infty }^{\infty }l\left\vert
l\right\rangle \left\langle l\right\vert .
\end{equation}%
This represents the simplest WSL system, which emerges from a uniform chain
possessing a single real energy ladder with spacing $\omega$, regardingless
of whether $\kappa $\ is real or imaginary.

(ii) In the case where $N=3$, the effective Hamiltonian is given by%
\begin{eqnarray}
H_{\mathrm{eff}}^{[3]} &=&-\kappa \sum_{j=-\infty }^{\infty }(\sqrt{3}%
|3j\rangle \langle 3j+1|+2|3j+1\rangle \langle 3j+2|  \notag \\
&&+\sqrt{3}|3j+2\rangle \langle 3j+3|+\mathrm{H.c.})  \notag \\
&&+\omega \sum_{l=-\infty }^{\infty }l\left\vert l\right\rangle \left\langle
l\right\vert .  \label{Heff_3}
\end{eqnarray}%
There are three sets of real ladders with spacing $3\omega $\ when $\kappa $%
\ is real. One of them contains a zero energy level, while all the levels of
other two ladders are symmetric about the zero energy.

On the other hand, when $\kappa $\ is imaginary, there are still three sets
of ladders with a spacing $3\omega $. The aforementioned constraints
necessitate the existence of zero energy. Numerical results show that the
other two are complex conjugates of each other.

(iii) In the case where $N=4$, the Hamiltonian is given by

\begin{eqnarray}
H_{\mathrm{eff}}^{[4]} &=&-\kappa \sum_{j=-\infty }^{\infty }(2|4j\rangle
\langle 4j+1|+\sqrt{6}|4j+1\rangle \langle 4j+2|\,  \notag \\
&&+\sqrt{6}|4j+2\rangle \langle 4j+3|\,+2|4j+3\rangle \langle 4j+4|  \notag
\\
&&+\mathrm{H.c.}) +\omega \sum_{l=-\infty }^{\infty }l\left\vert
l\right\rangle \left\langle l\right\vert .  \label{Heff_4}
\end{eqnarray}%
Numerical simulation shows that, when $\kappa $\ is imaginary, there are two
sets of real ladders and two sets of complex ladders. The two real ladders
form a new ladder with energy levels given by $\pm 2m\omega$,\ where $m=0$, $%
1$, .... The real parts of two complex ladders is given by $(2\pm 4m)\omega$%
,\ where $m=0$, $1$, ....

These results lead to the conclusion that for small $N$, in the case with
imaginary $\kappa $, there exists at least one set of real energy ladder.
This suggests that normal Bloch oscillations can be observed in such
non-Hermitian systems for a carefully chosen initial state. In contrast, an
arbitrary initial state should not exhibit a regular Bloch oscillation due
to the imaginary parts of the energy levels. However, the component of the
evolved state that involves the ladder with the maximal imaginary part
should dominate after a period of time. To demonstrate these points,
numerical simulations are performed for the dynamics driven by two typical
effective Hamiltonians. We consider the time evolution of an initial state $%
\left\vert \Psi \left( 0\right) \right\rangle $, which is a site state
corresponding to the multi-boson state, given by $\frac{1}{\sqrt{N!}}\left(
b_{j}^{\dagger }\right) ^{N}|\mathrm{vac}\rangle $. To characterize the
profile of the evolved state, we compute the Dirac probability distribution,
defined as

\begin{equation}
p_{l}(t)=\left\vert \left\langle l\right\vert \Psi \left( t\right) \rangle
\right\vert ^{2}=\left\vert \left\langle l\right\vert e^{-i\Lambda _{\mathrm{%
\max }}^{[N]}t}e^{-iH_{\mathrm{eff}}^{[N]}t}\left\vert \Psi \left( 0\right)
\right\rangle \right\vert ^{2}.  \label{p_l}
\end{equation}%
Here the prefactor is given by $\Lambda _{\mathrm{\max }}^{[N]}=$max$\left[ 
\func{Im}(E^{[N]})\right] $, representing the maximum of the imaginary part
of all the energy levels $\left\{ E^{[N]}\right\} $\ of the Hamiltonian $H_{%
\mathrm{eff}}^{[N]}$. This prefactor is introduced to reduce the exponential
growth of the probability in order to present a complete profile of the
evolved state. We plot $p_{l}(t)$ in Fig. \ref{fig:fig1}(a1, a2) and Fig. %
\ref{fig:fig2}(a1, a2), for the cases with $N=3$ and $N=4$, respectively.
These numerical results agree with our above analysis: (i) when taking real $%
\kappa $, the dynamics are quasi periodic; (ii) when taking imaginary $%
\kappa $, the dynamics becomes periodic after a period of time.

Before we conclude this section, we would like to point out that the
aforementioned results are obtained from effective Hamiltonians, which
describe non-interacting systems and subspaces of correlated bosons in the
large $U$ limit.

\section{Dynamics of correlated bosons}

\label{Dynamics of correlated bosons}

{ {According to the theorem, WSLs exist in a Bose-Hubbard model with
arbitrary parameters }$\kappa ${, }$U${\ and }$V${.
However, the numerical results presented so far have been essentially for
single-particle systems. In this section, we will focus on both the
Hermitian and non-Hermitian extended Bose-Hubbard models with moderate
values of }$U${\ and }$V${. We will provide numerical results
to show the existence of WSLs and present numerical results on the dynamics
of correlated bosons, which offer evident demonstrations of our rigorous
results.}}

 The feature of WSLs can be characterized by the quantity $%
\varepsilon $, which is defined as the sum of the real part of $E$ and the
imaginary part of $E$,%
\begin{equation}
\varepsilon =\func{Re}(E)+\func{Im}(E).  \label{dressed energy}
\end{equation}%
This quantity is referred to as dressed energy. Although $\varepsilon $\
does not have an evident physical meaning, it is still expressed in what is
known as the ladder form

\begin{equation}
\varepsilon =\varepsilon _{0}\pm m\omega ,
\end{equation}%
where $m=0$, $1$, $2$, .... The advantage of $\varepsilon $\ is that one can
demonsrate the complex ladder using the real ladder of $\varepsilon $. When
an exceptional point appears in the complex region of $E$, we can directly
observe the coalescence of two sets of $\varepsilon $\ ladder. It is
remarkable that the $\varepsilon $ ladders also exhibit level repulsion
around the exceptional points. To demonstrate these features we compute the
quantity $\varepsilon $\ as a function of $\kappa $, which is obtained by
the exact diagonalization of the Hamiltonian $H_{\mathrm{B}}$\ on finite
size chains with $N=3$ and $4$. In Fig. \ref{fig:fig3}, quantities $%
\varepsilon (\kappa )$ are plotted for both real and imaginary $\kappa $. As
expected, the plots of $\varepsilon (\kappa )$ are evidently several
families of curves. A coalescing ladder appears in the imaginary region of $%
\kappa $.

Based on these observations, one can detect the existence of such ladder
structure by the dynamical behavior of a given state. To demonstrate these
points, numerical simulations are performed for the dynamics driven by the
original Hamiltonian. We consider the time evolution of an initial state $%
\left\vert \Psi \left( 0\right) \right\rangle =\frac{1}{\sqrt{N!}}\left(
b_{j}^{\dagger }\right) ^{N}|\mathrm{vac}\rangle $, which is multi-boson
state. To characterize the profile of the evolved state, we compute the
Dirac probability distribution, defined as\ 
\begin{equation}
P_{j}(t)=\left\langle \Psi \left( t\right) \right\vert a_{j}^{\dag
}a_{j}\left\vert \Psi \left( t\right) \right\rangle =\left\vert a_{j}e^{-iH_{%
\text{B}}t}e^{-\Lambda _{\mathrm{\max }}t}\left\vert \Psi \left( 0\right)
\right\rangle \right\vert ^{2}  \label{Pj(t)}
\end{equation}%
for several cases and plot them in Fig. \ref{fig:fig1}(b1, b2) and Fig. \ref%
{fig:fig2}(b1, b2), for the cases with $N=3$ and $N=4$, respectively. Here
the prefactor is given by $\Lambda _{\mathrm{\max }}=$max$\left[ \func{Im}(E)%
\right] $, representing the maximum of the imaginary part of all the energy
levels $\left\{ E\right\} $\ of the Hamiltonian $H_{\mathrm{B}}$. Similarly,
this prefactor is introduced to reduce the exponential growth of the
probability in order to present a complete profile of the evolved state.

As predicted, we observe (i) Bloch-Zener oscillations for Hermitian systems;
(ii) stable oscillations after a period of time for non-Hermitian systems.
This indicates that the dynamics for the Hamiltonian $H_{\mathrm{B}}$\ with
moderate $U$ is similar to that for the\ effective Hamiltonian $H_{\mathrm{%
eff}}^{[N]}$. Such behavior provides a dynamic signature indicative of the
existence of the WSLs in the original Bose-Hubbard model.

\section{Summary}

\label{Summary}

In summary, we have developed a theory for a class of Hermitian and
non-Hermitian Hamiltonians that possess WSLs. Such Hamiltonians consist of
two distinct parts: an arbitrary Hamiltonian exhibiting translational
symmetry and a linear potential. {{The results are significant for
few-body systems or for {many-body} systems with few-body clusters.}
As examples, we have investigate the extended Bose-Hubbard model with
several bosons. Analytic and numerical approaches verified our conclusion in
the cases with large and moderate $U=V$, real and imaginary $\kappa $.} We
also demonstrated the application of the theorem by analyzing dynamics
driven by the Hubbard model using numerical simulations. Our findings
provide a method for analyzing the energy-level structure of general systems
and are expected to be both necessary and insightful for quantum engineering
in { quantum interacting} systems.

\acknowledgments This work was supported by National Natural Science
Foundation of China (under Grant No. 12374461).

%
\end{document}